\documentclass[12pt]{article}
\usepackage{epsfig}
\usepackage{axodraw}
\usepackage{epsfig}                            
\usepackage{graphicx}
\usepackage{rotate}
\usepackage{latexsym}
\newcommand\pubnumber{}
\newcommand\pubdate{\today}
\newcommand\hepnumber{hep-ph/0101299}

\def\csumb{Dipartimento di Fisica Teorica, Universit\`a di Torino, Italy\\
INFN, Sezione di Torino, Italy}
\def\support{\footnote{Work supported by the
European Union under contract HPRN-CT-2000-00149.}} 

\def\Title#1{\begin{center} {\Large\bf #1 } \end{center}}
\def\Author#1{\begin{center}{ \sc #1} \end{center}}
\def\Address#1{\begin{center}{ \it #1} \end{center}}

\newcommand\pubblock{\rightline{\begin{tabular}{l} \pubnumber\\
         \pubdate\\ \hepnumber \end{tabular}}}
\newenvironment{Abstract}{\begin{quotation}  }{\end{quotation}}
\newenvironment{Presented}{\begin{quotation} \begin{center} 
             Presented at the\end{center}
      \begin{center}\begin{large}}{\end{large}\end{center} \end{quotation}}

\makeatletter
\def\section{\@startsection{section}{0}{\z@}{5.5ex plus .5ex minus
 1.5ex}{2.3ex plus .2ex}{\large\bf}}
\def\subsection{\@startsection{subsection}{1}{\z@}{3.5ex plus .5ex minus
 1.5ex}{1.3ex plus .2ex}{\normalsize\bf}}
\def\subsubsection{\@startsection{subsubsection}{2}{\z@}{-3.5ex plus
-1ex minus  -.2ex}{2.3ex plus .2ex}{\normalsize\sl}}

\renewcommand{\@makecaption}[2]{%
   \vskip 10pt
   \setbox\@tempboxa\hbox{\small #1: #2}
   \ifdim \wd\@tempboxa >\hsize     
       \small #1: #2\par          
     \else                        
       \hbox to\hsize{\hfil\box\@tempboxa\hfil}
   \fi}

 \def\citenum#1{{\def\@cite##1##2{##1}\cite{#1}}}
\def\citea#1{\@cite{#1}{}}
 
\newcount\@tempcntc
\def\@citex[#1]#2{\if@filesw\immediate\write\@auxout{\string\citation{#2}}\fi
  \@tempcnta\z@\@tempcntb\m@ne\def\@citea{}\@cite{\@for\@citeb:=#2\do
    {\@ifundefined
       {b@\@citeb}{\@citeo\@tempcntb\m@ne\@citea\def\@citea{,}{\bf ?}\@warning
       {Citation `\@citeb' on page \thepage \space undefined}}%
    {\setbox\z@\hbox{\global\@tempcntc0\csname b@\@citeb\endcsname\relax}%
     \ifnum\@tempcntc=\z@ \@citeo\@tempcntb\m@ne
       \@citea\def\@citea{,}\hbox{\csname b@\@citeb\endcsname}%
     \else
      \advance\@tempcntb\@ne
      \ifnum\@tempcntb=\@tempcntc
      \else\advance\@tempcntb\m@ne\@citeo
      \@tempcnta\@tempcntc\@tempcntb\@tempcntc\fi\fi}}\@citeo}{#1}}
\def\@citeo{\ifnum\@tempcnta>\@tempcntb\else\@citea\def\@citea{,}%
  \ifnum\@tempcnta=\@tempcntb\the\@tempcnta\else
  {\advance\@tempcnta\@ne\ifnum\@tempcnta=\@tempcntb \else\def\@citea{--}\fi
    \advance\@tempcnta\m@ne\the\@tempcnta\@citea\the\@tempcntb}\fi\fi}
\makeatother

\newcommand{\spro}[2]{{#1}\cdot{#2}}
\newcommand{\ds}{\displaystyle}
\newcommand{\Ddrh}{{\ds\frac{1}{\hat{\varepsilon}}}}

\newcommand{\bec}{\begin{center}}
\newcommand{\eec}{\end{center}}
\newcommand{\vj}[4]{{\sl #1~}{\bf #2 }\ifnum#3<100 (19#3) \else (#3) \fi #4}
\newcommand{\ej}[3]{{\bf #1~}\ifnum#2<100 (19#2) \else (#2) \fi #3}

\newcommand{\vb}{V}

\newcommand{\ben}{\begin{enumerate}}
\newcommand{\een}{\end{enumerate}}
\newcommand{\asums}[1]{\sum_{#1}}
\newcommand{\bei}{\begin{itemize}}
\newcommand{\eei}{\end{itemize}}

\newcommand{\ib}{i}
\newcommand{\fe}{e}
\newcommand{\ff}{f}

\newcommand{\bqas}{\begin{eqnarray*}}
\newcommand{\eqas}{\end{eqnarray*}}
                     
\newcommand{\gz}{\Gamma_{_{\zb}}}

\newcommand{\me}{m_e}

\newcommand{\smans}{s^2}

\newcommand{\Reb}{{\rm{Re}}}

\newcommand{\bq}{\begin{equation}}                   
\newcommand{\eq}{\end{equation}}
\newcommand{\bqa}{\begin{eqnarray}}
\newcommand{\eqa}{\end{eqnarray}}
\newcommand{\ba}[1]{\begin{array}{#1}}
\newcommand{\ea}{\end{array}}
\newcommand{\lpar}{\left(}                            
\newcommand{\rpar}{\right)} 
\newcommand{\lrbr}{\left[}
\newcommand{\rrbr}{\right]}
\newcommand{\lcbr}{\left\{}
\newcommand{\rcbr}{\right\}} 
\newcommand{\ph}{\gamma}
\newcommand{\zb}{Z}
\newcommand{\wb}{W}

\newcommand{\barf}{\overline f}

\newcommand{\barb}{\overline b}

\newcommand{\mz}{M_{_Z}}
\newcommand{\mzs}{M^2_{_Z}}
\newcommand{\mzc}{M^3_{_Z}}

\newcommand{\mes}{m^2_e}

\newcommand{\gf}{G_{\ssF}}
\newcommand{\ssZ}{{\scriptscriptstyle{\zb}}}

\newcommand{\ssV}{{\scriptscriptstyle{\vb}}}
\newcommand{\ssA}{{\scriptscriptstyle{A}}}

\newcommand{\ssF}{{\scriptscriptstyle{F}}}

\newcommand{\ssL}{{\scriptscriptstyle{L}}}

\newcommand{\ssQ}{{\scriptscriptstyle{Q}}}

\newcommand{\ord}[1]{{\cal O}\lpar#1\rpar}
\newcommand{\als}{\alpha_{_S}}
\def\beq{\begin{equation}}
\def\eeq{\end{equation}}
\def\beqar{\begin{eqnarray}}
\def\eeqar{\end{eqnarray}}
\def\barr#1{\begin{array}{#1}}
\def\earr{\end{array}}
\def\bfi{\begin{figure}}
\def\efi{\end{figure}}
\def\btab{\begin{table}}
\def\etab{\end{table}}
\def\bce{\begin{center}}
\def\ece{\end{center}}

\def\nl{\nonumber\\}

\newcommand{\MeV}{\unskip\,\mathrm{MeV}}

\def\mathswitchr#1{\relax\ifmmode{\mathrm{#1}}\else$\mathrm{#1}$\fi}

\def\mathswitch#1{\relax\ifmmode#1\else$#1$\fi}

\def\cf{cf.\ }

\newcommand{\bfig}{\begin{center}\begin{picture}}
\newcommand{\efig}[1]{\end{picture}\\{\small #1}\end{center}}

\newcommand{\bmip}[2]{\begin{minipage}[t]{#1pt}\bfig(#1,#2)}
\newcommand{\emip}[1]{\efig{#1}\end{minipage}}

\newcommand{\beanon}{\begin{eqnarray*}}
\newcommand{\eeanon}{\end{eqnarray*}}

\renewcommand{\to}{\rightarrow}

\def\alf1{ {\alpha\over\pi} }
\newcommand{\rZf}{\rho^{\ff}_{_\zb}}
\newcommand{\gadu}[1]{\gamma_{#1}}
\newcommand{\tcif}{I^{(3)}_{\ff}}
\newcommand{\gdp}{\gamma_+}

\newcommand{\qf}{Q_f}
\newcommand{\kZdf}[1]{\kappa^{#1}_{_\zb}}

\newcommand{\Gvf}{{\cal{G}}^{\ff}_{_{V}}}
\newcommand{\Gaf}{{\cal{G}}^{\ff}_{_{A}}}

\newcommand{\gfd}{\gamma_5}                    
\newcommand{\gff}{\Gamma_{\ff}}
\newcommand{\fl}{l}
\newcommand{\fq}{q}
\newcommand{\mf}{m_f}
\newcommand{\srt}{\sqrt{2}}

\newcommand{\had}{{h}}

\newcommand{\gel}{\Gamma_{\fe}}
\newcommand{\gh}{\Gamma_{_{h}}}
\newcommand{\gzs}{\Gamma^2_{_{\zb}}}

\newcommand{\seffsf}[1]{\sin^2\theta^{#1}_{\rm{eff}}}

\newcommand{\gvf}{g^{\ff}_{_{V}}}
\newcommand{\gaf}{g^{\ff}_{_{A}}}
\newcommand{\afba}[1]{A^{#1}_{_{\rm FB}}}
\newcommand{\gl}{\Gamma_{\fl}}
\newcommand{\intfx}[1]{\int_{\scriptstyle 0}^{\scriptstyle 1}\,d#1}
\newcommand{\intfxy}[2]{\int_{\scriptstyle 0}^{\scriptstyle 1}\,d#1\,
                        \int_{\scriptstyle 0}^{\scriptstyle #1}\,d#2}
\newcommand{\cff}[7]{C_{#1}\lpar #2,#3,#4;#5,#6,#7\rpar}

\begin{document}
\begin{titlepage}
\pubblock

\vfill
\def\thefootnote{\fnsymbol{footnote}}
\Title{Precision Physics Near LEP Shutdown \\[5pt] and 
Evolutionary Developments}
\vfill
\Author{Giampiero Passarino\support}
\Address{\csumb}
\vfill
\begin{Abstract}
The most celebrated aspects of precision physics are briefly summarized.
New ideas are also introduced that may lead to further developments in the 
field of radiative corrections, with special emphasis to multi-loop 
numerical evaluation and to a correct treatment of QED radiation for
arbitrary processes.
\end{Abstract}
\vfill
\begin{Presented}
50 Years of Electroweak Physics \\[4pt]
A symposium in honor of Professor \\
Alberto Sirlin's 70th Birthday\\
October 27-28, 2000
\end{Presented}
\vfill
\end{titlepage}
\def\thefootnote{\arabic{footnote}}
\setcounter{footnote}{0}

\section{The experimental progress}

The experimental progress made in the 1990's is remarkable in 
many ways. With the right experiments we can reveal the fine details of 
high energy phenomena.
On the eve of LEP shutdown we known that physics at the 
microscopic scale has a basis as rationale as chemistry,
everything moves, not by magic, but because there is a mechanism at work.
Everything is encouraging for the major unsolved aspects of the breaking of 
electroweak gauge symmetry.

The blue band, giving the theoretical uncertainty in the $\chi^2$-distribution
as a function of the Higgs boson mass is, perhaps, the most celebrated figure 
of the whole LEP era.
\begin{figure}[t]
\begin{minipage}[t]{14cm}
{\begin{center}
\vspace*{-1.5cm}
\hspace*{-1.0cm}
\mbox{\epsfysize=10cm\epsfxsize=8cm\epsffile{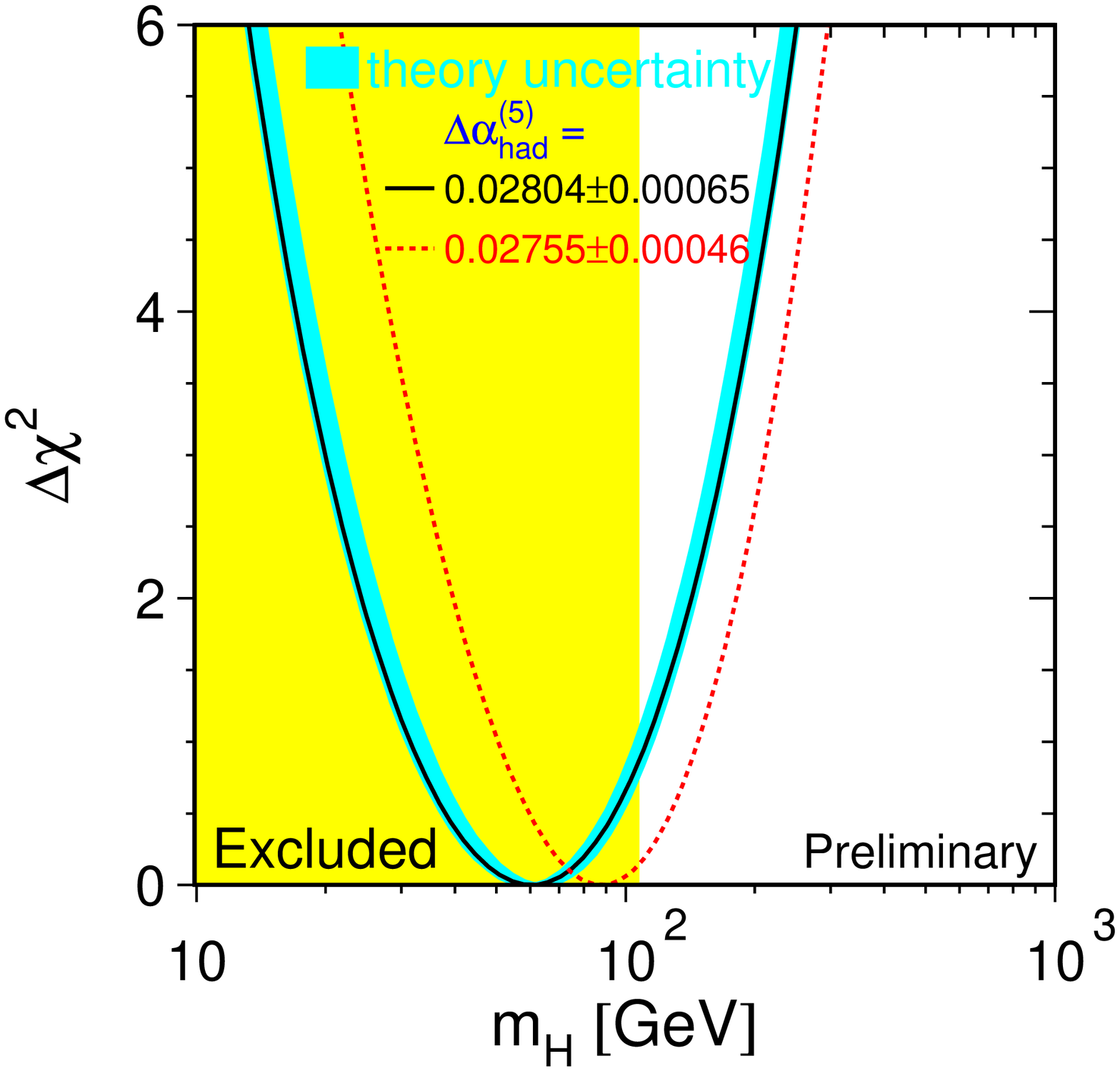}}
\end{center}}
\end{minipage}
\end{figure}
It is interesting to examine the environment in which the technology for 
producing this figure has been developed. One way is through the K\"uhn tower.
According to this dantescan representation we find at the top of the tower
a group of highly-specialized experts: QED, QCD etc.
Next to them we will find the Smugglers: those who assemble a
code mooncursing it.
At the lower level the Users are represented: the experimental community at 
large.

The Rosetta stone that keeps this world running is based on some sort of 
universal language that has been created to allow for a meaningful 
communication, the Pseudo-Observable language. 

\subsection{The language of Pseudo-Observables}

The PO are related to measured cross-sections and
asymmetries by some deconvolution or unfolding procedure and
the concept itself of pseudo-observability is rather difficult to define.
One way to introduce it is to say that the experiments 
measure some primordial (basically 
cross-sections and thereby asymmetries also) quantities which 
are then reduced to secondary quantities under
some set of specific assumptions. 
Within these assumptions the secondary
quantities, the PO, also deserve the label of observability.

\begin{figure}[t]
\begin{minipage}[t]{14cm}
{\begin{center}
\vspace*{-1.5cm}
\hspace*{-1.0cm}
\mbox{\epsfysize=10cm\epsfxsize=8cm\epsffile{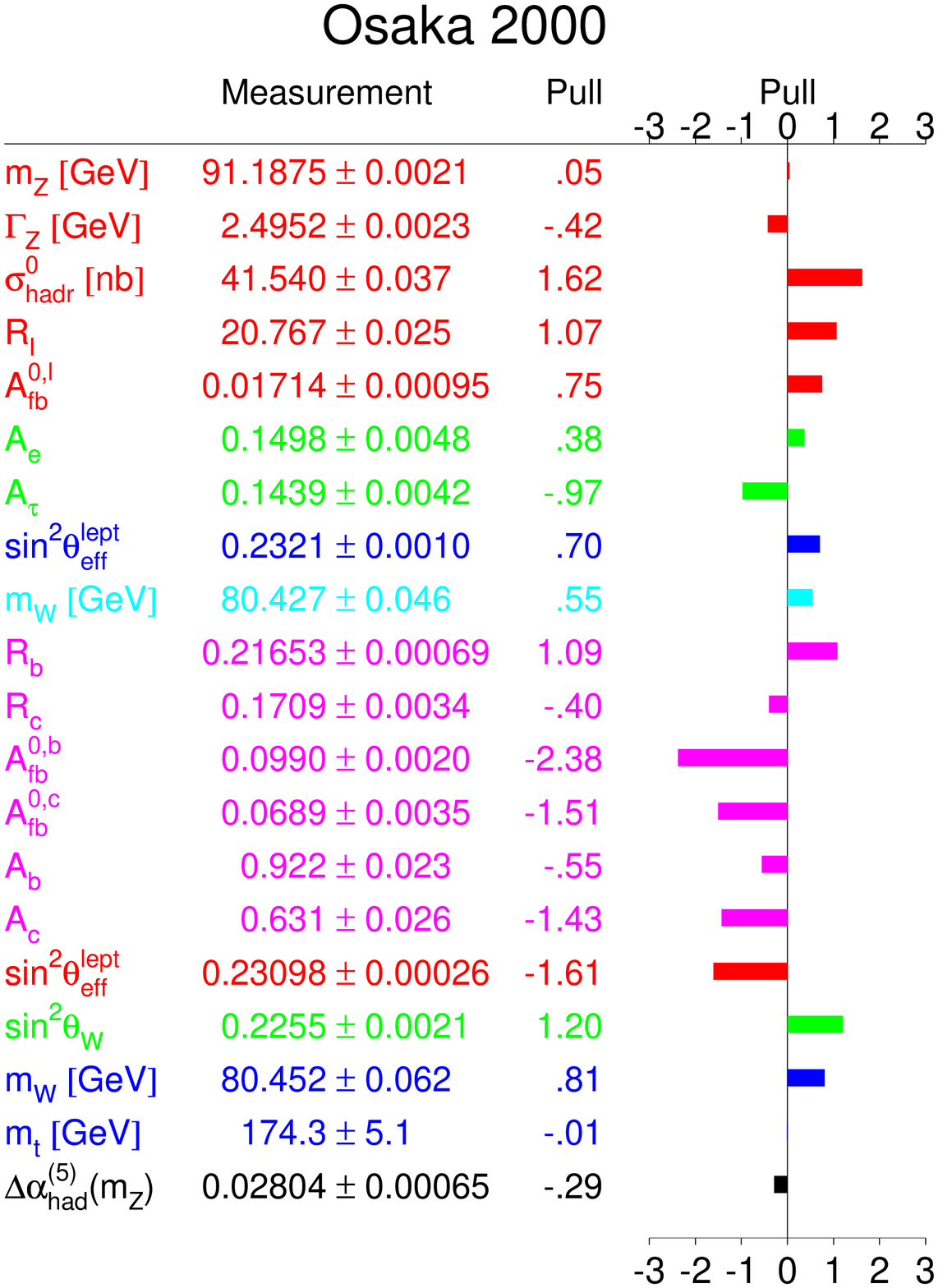}}
\end{center}}
\end{minipage}
\end{figure}
A quick example of this language is as follows.
The explicit formulae for the $\zb\ff\barf$ vertex are always 
written starting from a Born-like form of 
a pre-factor $\times$ fermionic current, 
where the Born parameters are promoted to effective,
scale-dependent parameters,
\bq
\rZf\gadu{\mu}\,\lrbr \lpar \tcif + \ib\,a_{\ssL}\rpar \gdp 
- 2\,\qf\kZdf{\ff} \smans + \ib\,a_{\ssQ}\rrbr
= \gadu{\mu}\,\bigl(\Gvf + \Gaf\,\gfd\bigr),
\label{prototype}
\eq
where $\gdp=1+\gfd$ and $a_{\ssQ,\ssL}$ are the SM imaginary 
parts.

By definition, the total and partial widths of the 
$\zb$ boson include also QED and QCD corrections.  
The partial decay width is therefore described by the following expression:
\bq
\gff\equiv\Gamma\lpar \zb\to\ff\barf\rpar = 4\,\cf\,\Gamma_0\,\bigl(
|\Gvf|^2\,R^{\ff}_{\ssV} + |\Gaf|^2\,R^{\ff}_{\ssA}\bigr) 
+\Delta_{_{\rm EW/QCD}}\;,
\label{defgammaf}
\eq
where $\cf = 1$ or $3$ for leptons or quarks $(\ff=\fl,\fq)$, and
$R^{\ff}_{\ssV}$ and $R^{\ff}_{\ssA}$
describe the final state 
QED and QCD corrections and take into account the fermion mass $\mf$.  
The last term,
\bq \Delta_{_{\rm EW/QCD}}= 
 \Gamma^{(2)}_{_{\rm EW/QCD}} - 
 \frac{\als}{\pi}\,\Gamma^{(1)}_{_{\rm EW}}\;, 
\eq 
accounts for the non-factorizable corrections.  
The standard partial width, $\Gamma_0$, is
\bq
\Gamma_0 = {{\gf\mzc}\over {24\srt\,\pi}} = 82.945(7)\,\MeV.
\eq
The peak hadronic and leptonic cross-sections are defined by
\bq
\sigma^0_{\had} = 12\pi\,\frac{\gel\gh}{\mzs\gzs} \qquad
\sigma^0_{\ell} = 12\pi\,\frac{\gel\gl}{\mzs\gzs} \;,
\eq
where $\gz$ is the total decay width of the $\zb$ boson, i.e, 
the sum of all partial decay widths.
The effective electroweak mixing angles 
( effective sinuses) are always defined by
\bq
4\,|\qf|\seffsf{\ff} = 1-\frac{\Reb\;\Gvf}{\Reb\;\Gaf} =
1-\frac{\gvf}{\gaf}, \quad \gvf = \Reb\;\Gvf, \qquad \gaf = \Reb\;\Gaf
\;.
\label{defeffsin}
\eq
The forward-backward asymmetry $\afba{}$ is defined via 
\bq
\afba{}=\frac{\sigma_{_{\rm{F}}}-\sigma_{_{\rm{B}}}}
             {\sigma_{_{\rm{F}}}+\sigma_{_{\rm{B}}}}\;,
\qquad
\sigma_{_{\rm{T}}}=\sigma_{_{\rm{F}}}+\sigma_{_{\rm{B}}}\;,
\eq
where $\sigma_{_{\rm{F}}}$ and $\sigma_{_{\rm{B}}}$
are the cross-sections for forward and backward scattering, respectively. 

\section{A new era}

Admittedly, we are entering a new era: 
we have been moving from a mediaeval architecture of radiative corrections 
to the renaissance of LEP period and now one should not 
indulge in deciphering the intimate meaning of the last digit in the 
last fit of the LEPEWWG. 

A new frontier is at the horizon: most likely it is goodbye to the 
one man show. Running a new RC project will be a little like running an 
experiment.

At the eve of LEP shutdown it is of some importance to summarize the present 
status of high precision physics~\cite{Bardin:1999gt}. 
For $e^+e^- \to \barf f$ all one-loop terms are known, including re-summation
of leading terms. At the two-loop level leading and next-to-leading terms
have been computed and included in codes like 
TOPAZ0~\citea{\citenum{Montagna:1993ai},\citenum{Montagna:1993py}}
and ZFITTER~\cite{Bardin:1999yd}.
For realistic observables initial state QED radiation is included via the 
structure function method, or equivalent ones. Final state QED is also 
available as well as the interference between initial and final 
states~\cite{Jadach:2000vf}.
Fine points in QED for $2\to2$ are as follows.
For $s$-channel all the $\ord{\alpha^2L^n},\, n=0,1,2\,\,$ terms are known
from explicit calculations, the  leading $\ord{\alpha^3L^3}$ is also available 
and they are important for the studies of the $\zb$ lineshape.

Differences and uncertainties amount to at most $\pm 0.1\,$MeV on
$\mz$ and $\Gamma_{\ssZ}$ and $\pm 0.01\%$ on 
$\sigma^0_{\rm h}$ (MIZA, TOPAZ0 and ZFITTER)~\cite{ewwg}
For non-annihilation processes (Bhabha) both structure-function
and parton-shower methods have been analyzed and the 
uncertainty is estimated to be $0.061\%$ from BHLUMI~\cite{Jadach:1997is}.
Certainly, full two-loop electroweak corrections are needed for GigaZ 
($10^9 \zb$ events) with a quest for a fast numerical evaluation of the 
relevant diagrams.

For $e^+e^- \to 4\,$fermions all tree-level processes are available and
$\ord{\alpha}$ electroweak corrections are known only for the $\wb\wb$-signal
and in double-pole approximation (DPA)~\cite{Denner:2000bj} 
and~\cite{Jadach:2000kw}.
$e^+e^- \to 4{\rm f} + \ph$ in Born approximation is also available for all 
processes~\cite{Grunewald:2000ju}. 
Dominant radiative corrections are inserted in single-$\wb$ production 
through the so-called Fermion-Loop 
scheme~\citea{\citenum{Beenakker:1997kn}-\citenum{Passarino:2000zh}}.

Fine points in QED for $2\to4$ are as follows
For $e^+e^- \to \wb\wb \to 4\,{\rm f}$ DPA gives the answer but,
for a generic process $e^+e^- \to 4\,{\rm f}$ QED radiation is included
by using $s$-channel structure-functions, \ i.e.
in leading-log approximation.
The latter are strictly applicable only if ISR can be separated unambiguously.
Otherwise their implementation may lead to an excess of radiation.
Preliminar investigations towards non--$s$ SF by GRACE and by 
SWAP~\cite{Grunewald:2000ju} gives an indication on how to implement
the bulk of the non-annihilation effect but still represent {\em ad hoc}
solutions. These methods, which are essentially based on a matching with the
soft photon emission, still contain an ambiguity on the energy scale 
selection with consequences on the predicted observables.

\section{Realistic observables}

How well do we know realistic observables in $e^+e^- \to 2\,{\rm f}$?
Pretty well around the $\zb$ peak, reasonably well up to LEP~2 energies, but 
fermion-pair corrections (PP) are not really under control.
Let us describe the structure of PP by starting with a simple case,
$e^+e^-$ PP-corrections to $e^+e^- \to \barb b$. 
PP is more than photon conversion, we will have the following sets of diagrams:
Multi-Peripheral, Initial State Singlet (ISS),
Initial State Non-Singlet (ISNS) and
Final State (FS).

Note that we include both $\ph$ and $\zb$ exchange.
On top of real pair production one has to include 
virtual $e^+e^-$ pairs. 
Let us compare as a typical example the sum of virtual and 
real pair corrections for primary hadrons and primary muons at 
$\sqrt(s)=189$~GeV.

The comparison between GENTLE, ZFITTER, and TOPAZ0 for
the diagram-based definition 
reveals maximum differences of $1.7 (1.5)$ per mill 
for inclusive hadrons (muons) and
$0.2 (0.4)$ per mill for high $s^\prime$ hadrons (muons).

However, PP are to be completed
expressly at higher energies and above vector boson thresholds.
The first evolutionary development is:
a 4f MC has to be fully interfaced with a 2f MC.

\section{QED radiation}

There are several 4f processes,
those with $t$-channel photons that are not dominated by annihilation.
Examples are: single-$\wb$ production and two-photon processes.
How to include the bulk of QED radiative corrections?
The second evolutionary development: is multi-photon radiation a one-scale or
a multi-scale convolution phenomenon?
\bqa
\sigma\lpar p_+p_- \to \{q_i\} + \mbox{QED}\rpar &\stackrel{?}{=}&
\int\,\prod_{\lambda=\pm} dx_{\lambda}\,D(x_{\lambda},?) \,
\sigma\lpar x_+p_+ x_-p_- \to \{q_i\}\rpar
\eqa
In the above equation the question mark means that the corresponding scale
has to be guessed. We need to understand how the standard SF-method is 
related to the exact YFS exponentiation.
In the standard YFS treatment of multiple photon emission we have
\bqa
\sigma\lpar p_++p_- \to \asums{i=1,2l}q_i + \asums{j=1,n}k_j\rpar &\sim& 
\int\,dPS_q \mid M_0\mid^2\,
E\lpar p_++p_--\asums{i}q_i\rpar,
\eqa
where $E$ is the spectral function defined through the usual eikonal factor:
\bqa
E(K) &=& \frac{1}{(2\,\pi)^4}\,\int\,d^4x\,\exp(i\spro{K}{x})\,E(x),  \nl
E(x) &=& \exp\lcbr \frac{\alpha}{2\,\pi^2}\,\int d^4k e^{i\spro{k}{x}}
\delta^+(k^2)\,\mid j^{\mu}(k)\mid^2\rcbr
\eqa
At this point we choose an alternative procedure were we do not separate the 
soft component from the hard one and compute some exact result valid for an 
arbitrary number of dimensions $n$ and for on-shell photons, i.e. $k^2=0$,
\bqa
I &=& \int d^nk\,e^{i\spro{k}{x}}\,{{\delta^+(k^2)}\over 
{\spro{p_i}{k}\,\spro{p_j}{k}}}
\eqa
In dimensional-regularization one has the following result, valid 
$\forall x^2$:
\bqa
I(x) &=& -\,\pi\,\rho\,\int_{_0}^{^1} \frac{du}{P^2}
\lpar \Ddrh + 2\,\ln 2 - \ln x^2 - \xi\,\ln\frac{\xi+1}{\xi-1}\rpar,  
\eqa
where we have defined a variable $\xi$ as the ratio
$\xi = |x_0|/r$, 
with an infinitesimal imaginary part attributed to $x_0$, i.e.
$x_0 \to x_0 + i\delta \, (\delta \to 0_+)$.
Furthermore, $P$ is the linear combination
$P = p_j + \lpar \rho p_i - p_j\rpar\,u$,
where we have defined $\rho$ to satisfy
$\lpar \rho p_i - p_j\rpar^2 = 0$, 
and $x_0, r$ are rewritten in covariant form as follows:
\bqa
x_0 = - {{\spro{P}{x}}\over {\sqrt{-P^2}}},  \quad
r^2 = x^2_0 + x^2.
\eqa
The last integral shows the infrared pole $\Ddrh$ and a collection of 
${\rm Li}_2$-functions. Therefore, $E(K)$ is not available in close form. 
The scheme that we want to propose defines a coplanar 
approximation~\cite{Chahine} to the exact spectral function,
\bqa
I^c_{ij} &\stackrel{\rm def}{=}& 
- \frac{2}{3}\,\pi \rho_{ij}\,{\cal F}_{\rm cp}\,
{1\over {p^2_j-\rho^2_{ij}\,p^2_i}}\,\ln\frac{\rho^2_{ij}p^2_i}{p^2_j}, \quad
I^c_{ii} \stackrel{\rm def}{=} 
- \frac{2}{3}\,\pi \rho_{ij}\,{\cal F}_{\rm cp}\,\frac{1}{m^2_i},
\nl
{\cal F}_{\rm cp} &=& \ln\lcbr e^{-\Delta_{\rm IR}} \,\, {{\spro{p_i}{x}\,
\spro{p_j}{x}}\over {m_im_j}}\rcbr,  \quad
\Delta_{\rm IR} = \Ddrh + {\rm constants}.
\eqa
Within the coplanar approximation we have
\bqa
E^{{\rm pair}\,<ij>}(K) &\stackrel{\rm cp}{\to}&  
\frac{1}{(2\,\pi)^2}\,\lcbr {{e^{-\Delta_{\rm IR}}}\over {m_im_j}}
\rcbr^{-\alpha A_{ij}}\,\frac{1}{\Gamma^2(\alpha A_{ij})}  \nl
{} &{}& {}  \nl
{} &\times& \int_0^{\infty}\, d\sigma d\sigma'\, 
\lpar \sigma\sigma'\rpar^{\alpha A_{ij} - 1}\, \delta^4\lpar
\sigma p_i + \sigma' p_j - K\rpar.
\eqa
This results explains why we have introduced the term {\em coplanar}.
Note that $\alpha A \sim \beta$ only when the corresponding 
invariant is much larger than mass${}^2$ but the above expression is valid 
for all regimes and it is easily generalized to $n$ emitters
with the result that~\footnote{A.Ballestrero, G.P. work in progress} 
in a process $2 \to n$ any external charged leg $i$ talks to all other 
charged legs, each time with a known scale $s_{ij}$ and with a known total 
weight proportional to
\bqa
x_i^{\alpha\,\lpar A^i_1 + \dots + A^i_I \rpar -1}
\,/\, \Gamma\lpar \alpha\,\lpar A^i_1 + \dots + A^i_I \rpar \rpar, 
\qquad 0 \le x_i \le 1
\eqa
Note that each $A$ has the appropriate sign, in/out, part/antp.
Furthermore, $I(i)$ is the number of pairs $<ij>$ with $i$ fixed.
The IR exponent is given by
\bqa
\alpha A &=& \frac{2\,\alpha}{\pi}\,\lcbr \frac{1+r^2}{1-r^2}\,\ln\frac{1}{r} -
1\rcbr,  \quad
\frac{\mes}{|t|} = \frac{r}{(1-r)^2}  
\eqa
For Bhabha scattering we will have the following combination:
\bqa
-A(s,\me)-A(t,\me)+A(u,\me) &=& 
\frac{2}{\pi}\,\Big[\ln\frac{st}{\mes u} - 1\Big],
\eqa
obtained as an exact result, not a guess.
\subsection{Conclusions for QED}

The structure-function language is still applicable but initial state 
structure functions evaluated for one scale is, quite obviously, not enough.
In any process each external leg brings one structure function;
since all charged legs talk to each other, each SF is not function of 
one {\em ad hoc} scale but all $<ij>$ scales enter into SF${}_i$.
The exact spectral-function is a convolution of SF
\bqa
E^{{\rm pair}\,<ij>}(K) &=& \int d^4K' \,\Phi(K')
E^{{\rm pair}\,<ij>}_{\rm cp}(K-K'),  \nl
\Phi(K) &=& \frac{1}{(2\,\pi)^4}\,\int d^4x\,\exp\lcbr i\,\spro{K}{x} +
\alpha\,\lpar I - I_{\rm cp}\rpar\rcbr  \,
= \delta(K)+\ord{\alpha}.
\eqa
Furthermore, IR-finite reminders and virtual parts can be added according to 
the standard approach of reorganizing the perturbative expansion.

\section{Multi-loop calculations}

The third evolutionary development can be summarized as follows.
A systematic approach to multi-loop calculations is needed.
Although a lot can be done by means of approximations
do we have a new, implementable algorithm for a full
scale attack to the problem?

There is a new idea around: the Tkachov theorem~\cite{fjodor}
or generalized Bernstein theorem (see also Bardin's talk).
\bqas
{\cal P}\lpar x,\partial\rpar \prod_i\,V_i^{\mu_i+1}(x) &=& B\,
\prod_i\,V_i^{\mu_i}(x).
\eqas
where ${\cal P}$ is a polynomial of $x = \lpar x_1 \cdots
x_n\rpar$ (a vector of Feynman parameters) and $\partial_{i} = 
\partial/\partial_i$; $B$ and all coefficients of
${\cal P}$ are polynomials of $\mu_i$ and of the
coefficients of $V_i(x)$.

The message is that we can do numerical calculations.
However, one-loop~\cite{Bardin:2000cf} versus multi-loop is exactly as
2d Ising model versus 3d Ising model.
For one-loop we have a universal master formula due to 
F.~V.~Tkachov~\cite{fjodor}:
\bqas
V(x) &=& x^t\,H\,x + 2\,K^t\,x + L,  \quad
{\cal P} = 1 - {{\lpar x+X\rpar\,\partial_x}\over {2\,\lpar\mu+1\rpar}},
\nl
X &=& K^t\,H^{-1}, \qquad B = L - K^t\,H^{-1}\,K.
\eqas
An example that everybody can do, the scalar $3$-point function
\bqas
C_0 &=& \intfxy{x}{y} V^{-1-\varepsilon/2}(x,y) = C^d_0+C^s_0+C^0_0,  \nl
V(x,y) &=& a\,x^2 + b\,y^2 + c\,xy + d\,x + e\,y + f -
i\,\varepsilon
\eqas
\bqas
C^d_0 &=&  2\,\intfxy{x}{y} V(x,y) \ln V(x,y) G_2^2 \Delta^{-2},  \nl
C^s_0 &=&
 \frac{1}{2}\,\intfx{x} \Bigl\{ V(x,x) \ln V(x,x)  
   \Bigl[ G_2 \Delta^{-2} (a_x-a_y)   
    + \frac{3}{2}\, \Delta^{-1} (a_x-a_y) 
     G_{11} \Delta_1^{-1}\Bigr]   \nl
{}&+&
  V(x,0) \ln V(x,0) \Bigl[ G_2 \Delta^{-2} a_y
  + \frac{3}{2}\, \Delta^{-1} a_y G_{12} \Delta_2^{-1}\Bigr]   \nl
{}&-&
  V(1,x) \ln V(1,x) \Bigl[ G_2 \Delta^{-2} a_x 
  + \frac{3}{2}\, G_2 \Delta^{-1} G_{13} \Delta_3^{-1}  
+ \frac{1}{2} G_2^2 \Delta^{-2}  
  + \frac{3}{2}\,  \Delta^{-1} a_x G_{13} \Delta_3^{-1}\Bigr]\Bigr\}, \nl
C^0_0 &=&  - \frac{1}{4}\, G_2 \Delta^{-1} a_{3x} \Delta_3^{-1} (b + c + e )
       - \frac{1}{8}\, G_2 \Delta^{-1} G_{13} \Delta_3^{-1} 
      (  \frac{4}{3} b + c + e ) \nl
{}&+&
 \frac{1}{12}\,G_2^2 \Delta^{-2}   ( 3 a + b + \frac{3}{2} c + 4 d 
   + 2 e + 6 f )
+ \frac{1}{4} \Delta^{-1} a_x a_{1x} \Delta_1^{-1}  ( a + b + c + d + e )\nl
{}&-&
\frac{1}{4}\Delta^{-1} a_x a_{3x} \Delta_3^{-1}   ( b + c + e )
       + \frac{1}{6}\,\Delta^{-1} a_x G_{11} \Delta_1^{-1}   ( a + b + c + 
        \frac{3}{4} d + \frac{3}{4} e )\nl
{}&-&
 \frac{1}{8}\, \Delta^{-1} a_x G_{13} \Delta_3^{-1}  ( \frac{4}{3} b
      + c + e )
       - \frac{1}{4}\, \Delta^{-1} a_y a_{1x} \Delta_1^{-1}  ( a + b + c + d
          e) \nl
{}&+&
 \frac{1}{4}\,\Delta^{-1} a_y a_{2x} \Delta_2^{-1}   ( a +  d )
- \frac{1}{6}\, \Delta^{-1} a_y G_{11} \Delta_1^{-1}   (  a + b +
       c + \frac{3}{4} d + \frac{3}{4} e ) \nl
{}&+&
 \frac{1}{6} \Delta^{-1} a_y G_{12} \Delta_2^{-1}   
         ( a + \frac{3}{4} d ) \nl
{}&+&
\frac{1}{4}\,V(0,0) \ln V(0,0) \Bigl[ \Delta^{-1} a_x a_{1x} \Delta_1^{-1} 
- \Delta^{-1} a_y a_{1x} \Delta_1^{-1}
       + \Delta^{-1} a_y a_{2x} \Delta_2^{-1}\Bigr]  \nl   
{}&-&
\frac{1}{4}\, V(1,0) \ln V(1,0) \Bigl[ G_2 \Delta^{-1} a_{3x} \Delta_3^{-1}
+ \Delta^{-1} a_x a_{3x} \Delta_3^{-1}
+4\, \Delta^{-1} a_y a_{2x} \Delta_2^{-1} \nl 
{}&+& \Delta^{-1} a_y G_{12} \Delta_2^{-1}\Bigr] \nl
{}&+&
\frac{1}{4}\,V(1,1) \ln V(1,1) \Bigl[ G_2 \Delta^{-1} a_{3x} \Delta_3^{-1}   
 + G_2 \Delta^{-1} G_{13} \Delta_3^{-1} 
- \Delta^{-1} a_x a_{1x} \Delta_1^{-1} \nl
&+& \Delta^{-1} a_x a_{3x} \Delta_3^{-1} \nl
{}&-&
 \Delta^{-1} a_x G_{11} \Delta_1^{-1} 
+ \Delta^{-1} a_x G_{13} \Delta_3^{-1} 
  \Delta^{-1} a_y a_{1x} \Delta_1^{-1}
   + \Delta^{-1} a_y G_{11} \Delta_1^{-1}\Bigr].
\eqas
Here, the Gram determinants are given by
\bqas
a_x &=& 2 d b-c e,  \;
a_y = 2 e a-d c,  \quad
a_{1x} = \frac{d+e}{2},  \;
a_{2x} = \frac{d}{2},  \;
a_{3x} = \frac{c+e}{2},  \nl
\Delta &=& f G_2-(b d^2-c d e+a e^2),  \nl
\Delta_1 &=& f G_{11} - \frac{(d+e)^2}{4},  \quad
\Delta_2 = f G_{12} - \frac{d^2}{4},  \quad
\Delta_3 = (a+d+f) G_{13} - \frac{(c+e)^2}{4},  \nl
G_2 &=& 4 a b-c^2,  \quad
G_{11} = a+b+c,  \quad
G_{12} = a,  \quad
G_{13} = b,  \nl
\eqas
Another example is given by the IR-divergent $C_0$- functions.
One can prove that $B = 0$ for
\bqas
a &=& b = f = m^2, \quad  \quad c = e = s - 2\,m^2, \quad d = -2\,m^2.
\eqas
As a consequence we have
\bqas
{}&{}&
\Bigl[1+P_x\,\frac{\partial}{\partial_x}+P_y\,\frac{\partial}{\partial_y}
\Bigr]\,V^{\mu+1}(x,y) = 0,  \quad
P_{x;y} = \frac{1-x;-y}{2(\mu+1)}. 
\eqas
This relation we use to write
\bqas
\intfxy{x}{y}\,V^{-1-\varepsilon/e}(x,y) &=& \frac{1}{\varepsilon}\,
\intfx{x}\,V^{-1-\varepsilon/e}(x,x),  \nl
\Bigl[ 1 + \frac{1}{2\,(\mu+1)}\,\lpar\frac{1}{2} - x\rpar\,\partial_x\Bigr]
V^{1+\mu}(x,x) &=& - \frac{1}{4}\,\lpar s - 4\,m^2\rpar\,V^{\mu}(x,x)
\eqas
\bqas
C^{\rm IR}_0 &=& \frac{1}{s-4 m^2} \Bigl\{ 
      \lpar - \frac{1}{2} + \frac{1}{\varepsilon} \rpar\,
      \Bigl[ 1 - \frac{1}{4}\,\ln V(0,0) - \frac{1}{4}\,\ln V(1,1)\Bigr]\nl
{}&+&
   \lpar - \frac{3}{4} + \frac{1}{2\,\varepsilon} \rpar
 \intfx{x} \ln V(x,x)\,\Bigr\}  
\eqas
Another example is the massless pentagon.
Here (but also for the massive one) one can prove that
\bqas
E_0 &=& 
\int_{\scriptstyle 0}^{\scriptstyle 1}\,dx_1\,
                        \int_{\scriptstyle 0}^{\scriptstyle 1-x_1}\,dx_2\,
                        \int_{\scriptstyle 0}^{\scriptstyle 1-x_1-x_2}\,dx_3\,
                        \int_{\scriptstyle 0}^{\scriptstyle 1-x_1-x_2-x_3}\,
                         dx_4  
\, \Bigl[x^t\,H^{-1}\,x + 2\,K^t\,x\Bigr]^{-3-\varepsilon/2},  \nl
\eqas
\[ H = \frac{1}{2}\,\left(\begin{array}{cccc}
0 & -s_{51} & s_{12}-s_{34} & s_{45} \\
- & -2\,s_{51} & -s_{34}-s_{51} & s_{23}-s_{51} \\
- & - & -2\,s_{34} & - s_{34} \\
- & - & - & 0
\end{array}\right)\]
\bqas
K_1 &=& 0, \quad K_2 = \frac{1}{2}\,s_{51}, 
\quad K_3 = \frac{1}{2}\,s_{34}, \quad K_4 = 0,  \quad
s_{ij} = - \lpar p_i+p_j\rpar^2.
\eqas
Now we raise the exponent and integrate by parts with
$B = 1/16\, s_{12} s_{23} s_{34} s_{45} s_{51}$.

In moving to multi-loop we recall that any diagram $G$ with $N_L$ legs and 
$n_l$ loops is representable as
\bqas
G &=& (-1)^N_L\,\lpar \frac{i}{{\pi}^{n/2}}\rpar^{n_l}\,
\Gamma\lpar N_L - \frac{n}{2}\,n_l\rpar\,\int\,
{{dz_G\,\delta\lpar 1-z_G\rpar}\over
{U^{n/2}\,\lpar V-i\,\varepsilon\rpar^{N_L-nn_l/2}}},  \nl
V &=& \asums{i} m^2_i\,z_i-\asums{i}\,q_i^2\,z_i - \frac{1}{U}\,
\asums{ij}\,B_{ij}\,\spro{q_i}{q_j}\,z_iz_j,  \nl
U &=& \asums{T}\prod_{z_i \in T}\,z_i = \mbox{det}\,\lpar U_{rs}\rpar, 
\quad U_{rs} = \asums{i} z_i\eta_{ir}\eta_{is}.
\eqas
$\eta_{is}$ is the projection of line 
$i$ along the loop $s$. $T$ is a co-tree.
Note that UV-singularities come from $U$ so that, for
finite diagrams, one should raise only the factor
\bqas
{\tilde V} &=& U\,\Bigl[\asums{i} m^2_i\,z_i-\asums{i}\,q_i^2\,z_i\Bigr] - 
\asums{ij}\,B_{ij}\,\spro{q_i}{q_j}\,z_iz_j.
\eqas
There are several comments.
For two-loops $U$ is quadratic and ${\tilde V}$ is cubic,
so that we have to construct a Bernstein functional relation
for a quintic or higher polynomial.
Try to do it naively in FORM and you will be told
`Input expansion buffer overflow. Try something else.'

For $N_L = 3,4$, we have ${\tilde V}$ to
a positive power and it is enough to regularize $G$ with some 
$K_S$ operation ($S\in G$).
For $N_L > 5$, we have ${\tilde V}$ to
a negative power, $U$ to a positive power (no UV
divergency), so we could raise a cubic.
For self-energies at $q^2 = 0$ it is easy.
Consider, for example, $n_l= 2,N_L= 3$, perform a 
projective transformation
\bqas
z_i &=& A_i\,u_i/\asums{j}A_j\,u_j, \qquad A_i = \frac{1}{m^2_i},
\eqas
to obtain ${\tilde V} = - U$ quadratic and,
\bqas
{\tilde V} &=&
       - x y   (  \frac{1}{m_1^2m_2^2} - \frac{1}{m_1^2m_3^2} + 
                  \frac{1}{m_2^2m_3^2} )
       - x   \frac{1}{m_1^2m_2^2}  \nl
{}&+& x^2 \frac{1}{m_1^2m_2^2}
       + y   ( \frac{1}{m_1^2m_2^2} - \frac{1}{m_1^2m_3^2} )
       + y^2   \frac{1}{m_2^2m_3^2}
\eqas
The crucial point being that we know how to raise a quadratic.
Finally, $N_L = 2\,(n_l+1)$ is nice, a single polynomial of degree
$n_l+1$.
Of course, vacuum diagrams are also easy. At two-loop level
we have
\bqas
G^{\rm vac}_2 &=& \lpar 4\,\pi\rpar^{\varepsilon-4}\,\Gamma\lpar 
\varepsilon-1\rpar\,\lpar K^tH^{-1}K\rpar^{-2}\,\intfxy{x}{y}  \nl
{}&\times& \Bigl[ 1 - {{\lpar x+K^tH^{-1}\rpar\partial}\over 
{\varepsilon-2}}\Bigr]\,
\Bigl[ 1 - {{\lpar x+K^tH^{-1}\rpar\partial}\over 
{\varepsilon}}\Bigr]\,U^{\varepsilon/2}(x,y),
\eqas
where $U$ has been presented above.
The same is, obviously true, for tadpoles.
What we are still missing is a real multi-loop example, 

Additional problems are represented by the fact that
${\tilde V}$ is an highly incomplete polynomial,
few non-zero coefficients.
${\cal P}$ can be found via a direct study of linear
systems of $n$ equations ($n \gg 100$) in $m$
variables and very often the matrix of coefficients has rank
$<$ n, so the system is generally impossible and,
we need to go to higher $n,m$, i.e. beyond
$n_{\rm min} = \min \{n<m\}$ (try one-loop master formulae with 
$K = L = 0$!) and your algebraic manipulation will exhaust 
your computer before you find a solution.

A simple counting shows the following situation: given
\bqas
{\cal P} &=& P_n + P^i_{n+1}\partial_i + P_{n+2}^{ij}\partial_i\partial_j +
\dots
\eqas
four variables and a cubic $V$ require 
$n_{\rm eq}= 126(330,715)$ and $n_{\rm var}= 155(415,871)$ for
$n= 2(1,0)$ and first(second, third) derivatives.
Simple examples already show that, for realistic 
polynomials $V$, one has to go beyond second order in derivatives.

The relation between $B = 0$ and Landau singularities is
still to be examined, although $B = 0$ contains all
singularities of $G$, physical and spurious ones.
$L-K^tH^{-1}K$ (or $B$ in general) is not a 
Gram determinant but its spurious zeros, if any, are a 
vexation, as much as those of Gram determinants in the standard reduction
procedure.
For $\cff{0}{-m_a^2}{-m_b^2}{-s}{m_b}{M}{m_a}$ we obtain
\bqas
G_{11} &=& -\frac{1}{4}\,\lambda\lpar s,m_a^2,m_b^2\rpar, \quad
G_{12} = -\frac{1}{4}\,\lambda\lpar M^2,m_a^2,m_b^2\rpar, \quad
G_{13} = -\frac{1}{4}\,\lambda\lpar M^2,m_a^2,m_b^2\rpar, \nl
G_2 &=&  \Bigl[ \lpar m_a^2-m_b^2\rpar^2 - M^2 s \Bigr] 
\, \Bigl[ s + M^2  - 2\,\lpar m_a^2 + m_b^2\rpar\Bigr].
\eqas

\end{document}